\documentclass{article}

\usepackage[nonatbib,preprint]{neurips_2024}

\usepackage[utf8]{inputenc} 
\usepackage[T1]{fontenc}    
\usepackage{hyperref}       
\usepackage{url}            
\usepackage{booktabs}       
\usepackage{amsfonts}       
\usepackage{nicefrac}       
\usepackage{microtype}      
\usepackage{xcolor}         
\usepackage{graphicx}
\usepackage{gensymb}
\usepackage[backend=biber,natbib=true,style=authoryear]{biblatex}

\title{Modulated Adaptive Fourier Neural Operators for Temporal Interpolation of Weather Forecasts}

\author{%
  Jussi Leinonen \\ 
  NVIDIA Corporation\\
  Santa Clara, CA 95051 \\
  \texttt{jleinonen@nvidia.com} \\
  \And
  Boris Bonev \\
  NVIDIA Corporation\\
  Santa Clara, CA 95051 \\
  \texttt{bbonev@nvidia.com} \\
  \AND
  Thorsten Kurth \\
  NVIDIA Corporation\\
  Santa Clara, CA 95051 \\
  \texttt{tkurth@nvidia.com} \\
  \And
  Yair Cohen \\
  NVIDIA Corporation\\
  Santa Clara, CA 95051 \\
  \texttt{yacohen@nvidia.com} \\
}

\begin{document}

\maketitle

\begin{abstract}
  Weather and climate data are often available at limited temporal resolution, either due to storage limitations, or in the case of weather forecast models based on deep learning, their inherently long time steps. The coarse temporal resolution makes it difficult to capture rapidly evolving weather events. To address this limitation, we introduce an interpolation model that reconstructs the atmospheric state between two points in time for which the state is known. The model makes use of a novel network layer that modifies the adaptive Fourier neural operator (AFNO), which has been previously used in weather prediction and other applications of machine learning to physics problems. The modulated AFNO (ModAFNO) layer takes an embedding, here computed from the interpolation target time, as an additional input and applies a learned shift--scale operation inside the AFNO layers to adapt them to the target time. Thus, one model can be used to produce all intermediate time steps. Trained to interpolate between two time steps $6$~h apart, the ModAFNO-based interpolation model produces $1$~h resolution intermediate time steps that are visually nearly indistinguishable from the actual corresponding $1$~h resolution data.  The model reduces the RMSE loss of reconstructing the intermediate steps by approximately $50\%$ compared to linear interpolation. We also demonstrate its ability to reproduce the statistics of extreme weather events such as hurricanes and heat waves better than $6$~h resolution data. The ModAFNO layer is generic and is expected to be applicable to other problems, including weather forecasting with tunable lead time.
\end{abstract}

\section{Introduction}

The urgent need to address climate change is driving an increasing demand for climate and weather data and more accurate predictions. Beyond meteorological services, such data are used by numerous sectors such as insurance, investing, energy, transport, infrastructure management and emergency responders. 

The demand for higher resolution, customized data products and probabilistic forecasts (typically achieved with \emph{ensemble} forecasts, i.e. executing a simulation multiple times from different initial conditions) is rapidly increasing the atmospheric data storage requirements, with the largest data archives approaching exabyte size \citep{ECMWF2024FactsFigures}. Such data volumes are only feasible to manage for the largest data centers, and even then at great cost.

Weather forecast models based on machine learning (ML) have recently achieved skill competitive with state-of-the-art numerical weather prediction (NWP) models while requiring orders of magnitude less time, energy and money to produce a forecast \citep{Pathak2022FourCastNet,Kurth2023FourCastNet,EbertUphoff2023Outlook,Lam2023GraphCast,Chen2023FengWu,Price2024Gencast,Zhong2024FuxiENS}. ML weather models make it plausible to avoid the long-term storage of simulation results, instead executing forecasts on demand and thus drastically requiring data storage costs. 

Most ML forecast models execute long time steps, for example $6$~h for GraphCast \citep{Lam2023GraphCast} and FourCastNet \citep{Pathak2022FourCastNet}. This can leave gaps in the data, especially when predicting rapidly evolving extreme weather events. PanguWeather \citep{Bi2023PanguWeather} consists of multiple models with different time steps ($1$~h, $3$~h, $6$~h and $24$~h) to achieve a $1$~h resolution, but this requires the training of multiple models and may introduce inconsistencies in the final simulation: For example, a $23$~h forecast is executed as $3 \times 6\ \mathrm{h} + 3\ \mathrm{h} + 2 \times 1\ \mathrm{h}$, which is not guaranteed to be consistent with following prediction, which is obtained from a single time step of the $24$~h model.

In this work, we assume that our input data has coarse temporal resolution, whether due to archiving constraints or as a result of using a long time step ML weather model, and treat the recovery of the intermediate time steps as an \emph{interpolation} rather than forecasting problem. Since deep neural networks have proven to be suitable for building powerful weather forecasting models, it can be expected that a network that learns to model the physics of the atmosphere should be able to produce a much better result than simple (e.g. linear) interpolation methods. Based on this premise, we introduce an adaptable architecture to predict arbitrary time steps using a single model. This is based on the adaptive Fourier neural operator (AFNO) used in the original FourCastNet model, with an additional time-dependent modulation in both the spatial and spectral domains to adjust the network to the desired time step.

\textbf{Contributions:} 
We introduce the \emph{modulated} adaptive Fourier neural operator (ModAFNO), which applies a scale--shift operation to spatial and spectral domain feed-forward neural networks to condition their output on the time step. The utility of the approach is demonstrated with a weather interpolation model, which uses ModAFNO to reconstruct global weather states at $1$-hourly resolution from $6$-hour resolution input data.

\section{Model}

The problem considered in this work is formulated as follows: given the global atmospheric states at times $t_0$ and $t_1$, as well as additional context variables and a target time $t$ with $t_0 \leq t \leq t_1$, predict the state at time $t$. 

\subsection{Architecture} \label{sect:architecture}

We base our model on the AFNO neural network architecture, which was used in the original FourCastNet model and has been shown to be adaptable to many physical problems \citep{Guibas2022AFNO,NVIDIA2023Darcy,Leinonen2023LDCast}. The AFNO implementation in NVIDIA Modulus (\url{https://github.com/NVIDIA/modulus}), available under the Apache-2.0 License, was used as the baseline. The ModAFNO architecture is also now available in Modulus.

The AFNO FourCastNet employs a patch encoding layer, a series of AFNO blocks, and a decoding layer. The AFNO blocks learn neural network operations both in the spatial domain and in the Fourier spectral domain. Formally, the $k$th AFNO block, operating on the output $x_{k-1}$ of the previous block, can represented as
\begin{eqnarray}
\mathrm{Filter}_k(x) &=& \mathrm{IDFT}(S_\lambda(\mathrm{MLP}^\mathrm{f}_k(\mathrm{DFT}(x)))) \label{eq:afno_1} \\
y_{k} &=& \mathrm{LayerNorm}^\mathrm{f}_k(\mathrm{Filter}_k(x_{k-1})) + x_{k-1} \label{eq:afno_2} \\
x_{k} &=& \mathrm{LayerNorm}^\mathrm{s}_k(\mathrm{MLP}^\mathrm{s}_k(y_{k})) + y_{k} \label{eq:afno_3}
\end{eqnarray}
In Eq.~\ref{eq:afno_1}, representing the spectral-domain filter, $\mathrm{DFT}$ and $\mathrm{IDFT}$ are the discrete 2D Fourier transform and the inverse 2D DFT, respectively, $S_\lambda(x) = \mathrm{sign}(x) \max(|x|-\lambda, 0)$ is a sparsity-promoting soft-thresholding operation, and $\mathrm{MLP}^\mathrm{f}_k$ is a feed-forward neural network (i.e. multilayer perceptron, MLP) operating in the spectral domain. Since the output of the DFT is complex-valued, $\mathrm{MLP}^\mathrm{f}_k$ is a complex-valued network. In Eq.~\ref{eq:afno_2}, the filter result is normalized with layer normalization (LayerNorm). In Eq.~\ref{eq:afno_3}, $\mathrm{MLP}^\mathrm{s}_k$ acts in the spatial domain and is a real-valued MLP that performs channel mixing with shared weights for each point; the result of this MLP is then passed through layer normalization. Residual connections are applied in both Eqs.~\ref{eq:afno_2} and~\ref{eq:afno_3}.

\begin{figure}
  \centering
  \includegraphics[width=0.85\linewidth]{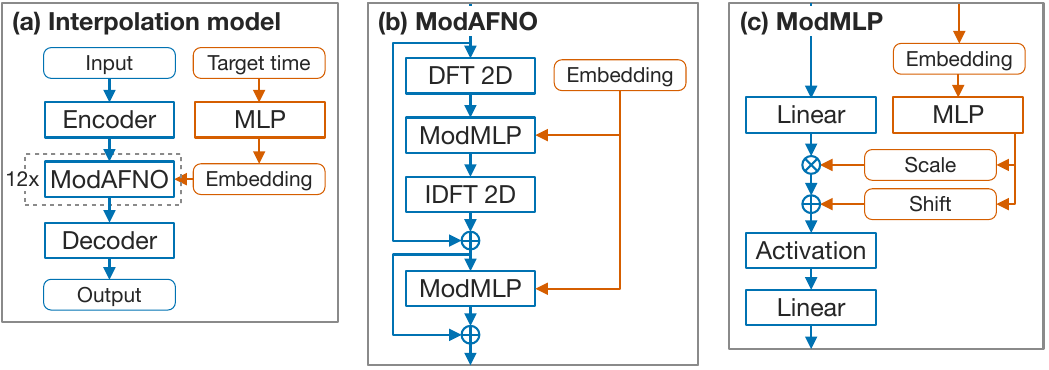}
  \caption{a) Overview of the ModAFNO interpolation network. b) The modulated AFNO layer. A time-dependent embedding is added to the MLPs in the spectral and spatial domains to enable temporal interpolation. c) Detailed view of the modulated MLP. A separate MLP produces the scale and shift activations from the time embedding.} \label{fig:modafno_network}
\end{figure}
To allow the AFNO to adapt to a desired time step, we introduce an embedding-dependent scale--shift operation to the AFNO layer in both the spatial and spectral stages, as illustrated in Fig.~\ref{fig:modafno_network}. This novel architectural feature is inspired by the use of a similar operation in convolutional networks in MetNet-2 \citep{Espeholt2022MetNet2} and MetNet-3 \citep{Andrychowicz2023MetNet3}. The resulting \emph{modulated} AFNO (ModAFNO) block can be expressed as
\begin{eqnarray}
\mathrm{ModFilter}_k(x,q) &=& \mathrm{IDFT}(S_\lambda(\mathrm{ModMLP}^\mathrm{f}_k(\mathrm{DFT}(x),q))) \label{eq:modafno_1} \\
y_{k} &=& \mathrm{LayerNorm}^\mathrm{f}_k(\mathrm{ModFilter}_k(x_{k-1},q)) + x_{k-1} \label{eq:modafno_2} \\
x_{k} &=& \mathrm{LayerNorm}^\mathrm{s}_k(\mathrm{ModMLP}^\mathrm{s}_k(y_{k},q)) + y_{k} \label{eq:modafno_3} .
\end{eqnarray}
The modulated MLPs (ModMLP) in the spectral (Eq.~\ref{eq:modafno_1}) and spatial (Eq.~\ref{eq:modafno_3}) domains are of the form
\begin{eqnarray}
    \mathrm{ModMLP}^{\mathrm{f}}_k(x,q) &=& \mathrm{Linear}^{\mathrm{f}}_{k,2}(\mathrm{ReLU}(\mathrm{Linear}^{\mathrm{f}}_{k,1}(x) \cdot w^\mathrm{f}_k(q) + b^\mathrm{f}_k(q))) \\
    \mathrm{ModMLP}^{\mathrm{s}}_k(x,q) &=& \mathrm{Linear}^{\mathrm{s}}_{k,2}(\mathrm{GELU}(\mathrm{Linear}^{\mathrm{s}}_{k,1}(x) \cdot w^\mathrm{s}_k(q) + b^\mathrm{s}_k(q)))
\end{eqnarray}
that is, they follow the typical flow of an MLP with linear layers combined with a nonlinear activation function, except a scale--shift operation is used to modulate the MLP in both the spectral and spatial domains. In both cases, the scale factors $w$ and the shifts $b$ only have channel and batch dimensions and are shared across all positions, adding only a small amount of computation. The spectral domain factors $w^\mathrm{f}$ and $b^\mathrm{f}$ are complex numbers, hence a complex multiplication and addition are performed in $\mathrm{ModMLP}^{\mathrm{f}}$, while the corresponding operations in $\mathrm{ModMLP}^{\mathrm{s}}$ are real-valued. The choice of activation functions follows the original AFNO architecture. The encoder and decoder architectures are also inherited unchanged from AFNO.

The embedding $q$ can, in principle, condition the network to many different types of inputs. In our model, we want the embedding to depend on the interpolation target time $t$; therefore we compute $q$ using a sinusoidal time embedding $\mathrm{EMB}$ that is passed through an MLP. Then, each of $w^\mathrm{f}_k(q)$, $b^\mathrm{f}_k(q)$, $w^\mathrm{s}_k(q)$, $b^\mathrm{s}_k(q)$ is computed from $q$ using a learned MLP that is unique to each stage $k$.
\begin{eqnarray}
    q &=& \mathrm{MLP}^q \left ( \mathrm{EMB} \left ( \frac{t-t_0}{t_1-t_0} \right ) \right ) \\
    w^{\{\mathrm{f},\mathrm{s}\}}_k(q) &=& \mathrm{MLP}^{w\{\mathrm{f},\mathrm{s}\}}_k(q) \\
    b^{\{\mathrm{f},\mathrm{s}\}}_k(q) &=& \mathrm{MLP}^{b\{\mathrm{f},\mathrm{s}\}}_k(q).
\end{eqnarray}

\subsection{Implementation and training} \label{sect:training}

\subsubsection{Data} \label{sect:data}

We used data from the global ECMWF Reanalysis v5 \citep[ERA5;][]{Hersbach2020ERA5} from the European Centre for Medium-Range Weather Forecasts (ECMWF) to train our model. This dataset gives an estimate of the historical state of the atmosphere at $1$~h temporal resolution on a $0.25\degree$ resolution latitude--longitude grid (with $721 \times 1440$ grid points per time step and variable). It is available under the open Licence to Use Copernicus Products at the Copernicus Climate Data Store \citep{Hersbach2023ERA5} or Google Cloud (\url{https://cloud.google.com/storage/docs/public-datasets/era5}). ERA5 is also the dataset used to train most global weather forecast models, including FourCastNet, PanguWeather and GraphCast.

The data downloaded from ERA5 for model training comprise pressure level height ($z$), temperature ($T$), specific humidity ($q$) as well as winds in the west--east ($u$) and north--south ($v$) directions. These are at $13$ pressure levels each ($50$, $100$, $150$, $200$, $250$, $300$, $400$, $500$, $600$, $700$, $850$ and $1000$~hPa). Furthermore, wind $u$/$v$ components at $10$~m and $100$~m heights from the surface, temperature at $2$~m height, surface pressure, mean sea level pressure and total column water vapor are included, giving a total of $73$ variables. The data were acquired at the full $1$~h temporal resolution and span the years $1980$--$2017$, with a total data volume of approximately $95$~TB.

To provide context information to the model, we add external forcings that are either constant or can be computed using known equations from time stamps. The constant inputs are the sine and cosine of latitude and longitude (which act as a positional embedding), land-sea mask and surface elevation; the single computed input is the cosine of the solar zenith angle.

\subsubsection{Implementation}

The model consists of the encoder, $12$ ModAFNO layers and the decoder. It processes the entire global ERA5 grid, except for the southernmost latitude since the model is limited to even-numbered dimensions. It thus has spatial input and output dimensions of $720 \times 1440$ (reduced from $721$ rows due to the removal of the southernmost latitude). The model has one tensor input, constructed by concatenating the states at $t_0$ and $t_1$ (normalized to zero mean and unit variance), the constant context variables, and the solar zenith angle at $t_0$, $t$ and $t_1$, giving a total of $155$ input channels ($2 \times 73$ atmospheric state variables and $9$ auxiliary inputs). It also has one scalar input containing the interpolation target time $t$, and a tensor output with $73$ channels.

Given the similarity of the AFNO and ModAFNO models, we initially trained followed the training setup of the AFNO FourCastNet. However, we found that this training configuration sometimes produced artifacts at the edges of the $8 \times 8$ tiles used by the encoder and decoder. We reduced the tile size to $2 \times 2$ and the number of channels to $512$, and removed the block-diagonal structure from the spectral part (effectively reducing the number of blocks to $1$) to stabilize training. leading to a model with fewer parameters but higher computational and memory requirements. This configuration produced a model with better skill than the original and resolved the artifact issue. Further reduction of the tile size to $1 \times 1$ produced a model with lower skill but much higher memory requirements; thus we did not explore this further. Other than the abovementioned hyperparameters, we used model and training configurations identical to the AFNO FourCastNet.

\subsubsection{Training}

We trained the model with $6$ million samples of training data, using the Adam optimizer \citep{Kingma2014Adam} starting with a learning rate of $5 \times 10^{-4}$ and decaying to $0$ over the training period using a cosine learning rate schedule. Mean square error weighted by the cosine of the latitude (i.e. proportional to the area of the grid points) was used as the training objective. The years $1980$--$2016$ were used for training while $2017$ was held back for testing. The training was performed on an NVIDIA DGX SuperPOD, using $64$ NVIDIA H100 GPUs and a global batch size of $64$, requiring approximately $1300$ GPU hours. GPU memory usage was approximately $72$~GB/GPU during training and $22$~GB/GPU during inference.

\section{Results}

To evaluate the effectiveness of our method, we evaluate the trained model using statistics of the reconstruction error, as well as a set of weather scenarios, which include Storm Ophelia, Hurricane Irma and the European heatwave of August $2017$. The method is compared to ground truth observations at $6$-hourly and $1$-hourly resolutions, as well as linear interpolation. Since we restrict our problem definition to reconstructing the weather between two known time steps, we do not consider interpolation methods that use more than two time steps to produce each output.

\subsection{Interpolated wind speeds during Storm Ophelia}

\begin{figure}
  \centering
  \includegraphics[width=\linewidth]{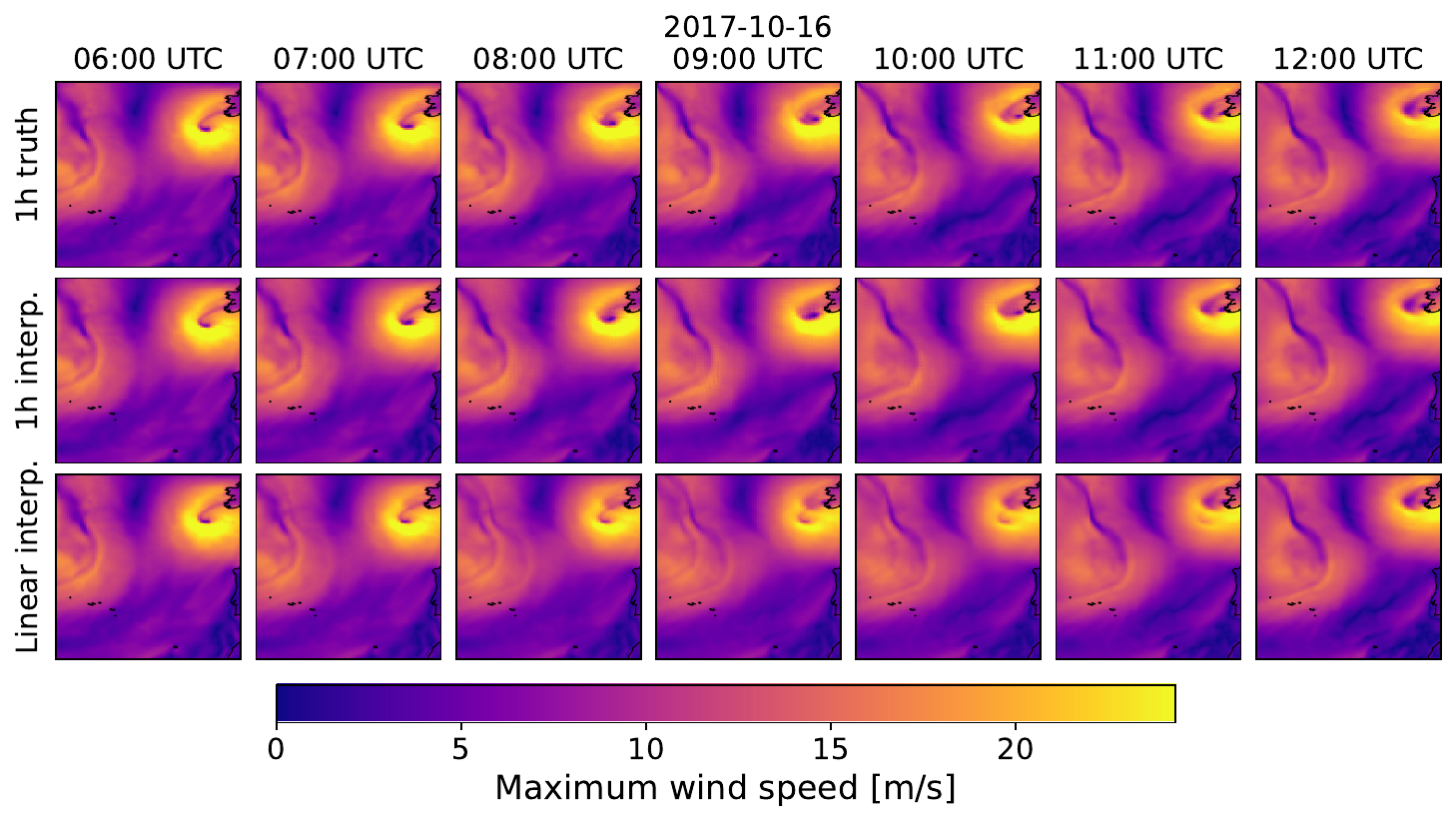}
  \caption{Example of ModAFNO interpolation of wind speeds from Storm Ophelia approaching Ireland on October 16, 2017. Top row: the actual ERA5 wind speed; middle row: the ModAFNO interpolated wind speed; bottom row: wind speed obtained with linear interpolation.} \label{fig:wind_case}
\end{figure}

We demonstrate results from the ModAFNO interpolation model in Fig.~\ref{fig:wind_case}. This example shows Storm Ophelia on the Atlantic Ocean approaching Ireland on October 16, 2017. The ModAFNO interpolation result is compared to the true $1$~h data and linear interpolation. The interpolation result is visually nearly indistinguishable from the true $1$~h data, correctly capturing the motion of the storm center on the top right and the front on the left side of the image. Meanwhile, the linear interpolation demonstrates the shortcomings of the $6$~h data by duplicating the fast-moving features in the middle frames, producing unrealistic weather states.

\subsection{Interpolation error}

\begin{figure}
  \centering
  \includegraphics[width=0.75\linewidth]{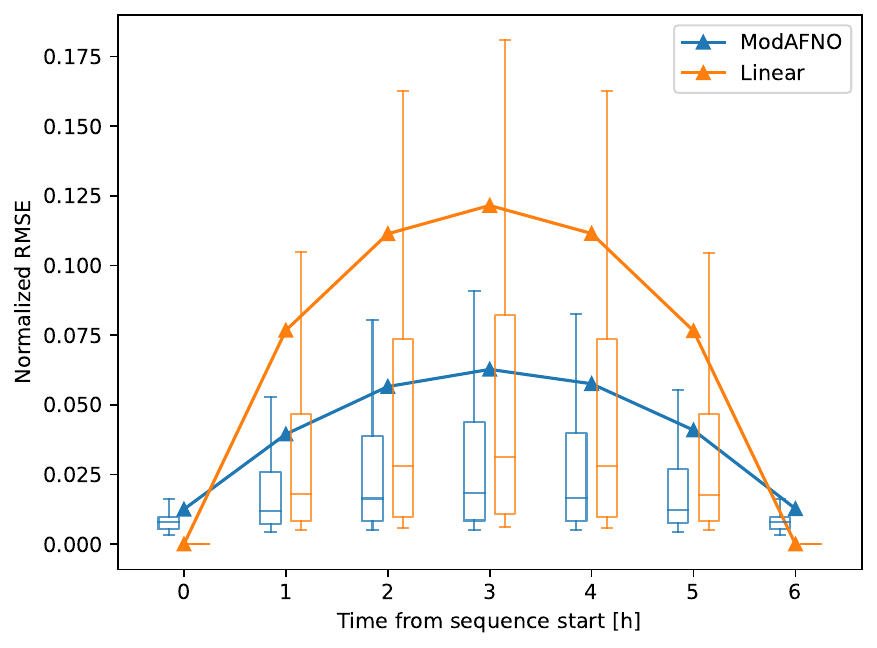}
  \caption{Normalized RMSE error of the interpolation for the ModAFNO model (blue) and linear interpolation (orange) as a function of the interpolation time. The line with triangles indicates the mean of the pointwise RMSE while the box plots indicate the median (midline of the box), the 25th and 75th percentiles (ends of the box) and the 9th and 91st percentiles (whiskers).} \label{fig:interp_error}
\end{figure}

In Fig.~\ref{fig:interp_error}, we show the statistics of the interpolation RMSE of the ModAFNO model against the linear interpolation baseline as a function of the interpolation time. The errors were computed from $1000$ random samples from the test dataset, using the values normalized to zero mean and unit variance to give approximately equal weight to each of the $73$ variables. The linear interpolation has zero error at the endpoints by definition, while the ModAFNO model has a small amount of error remaining at the endpoints due to the lossy nature of its encoding and decoding. At the other interpolation times, ModAFNO achieves approximately $50\%$ reduction in RMSE compared to the linear interpolation. The errors of both models display highly skewed statistics.

\subsection{Maximum wind speeds in Hurricane Irma}

\begin{figure}
  \centering
  \includegraphics[width=\linewidth]{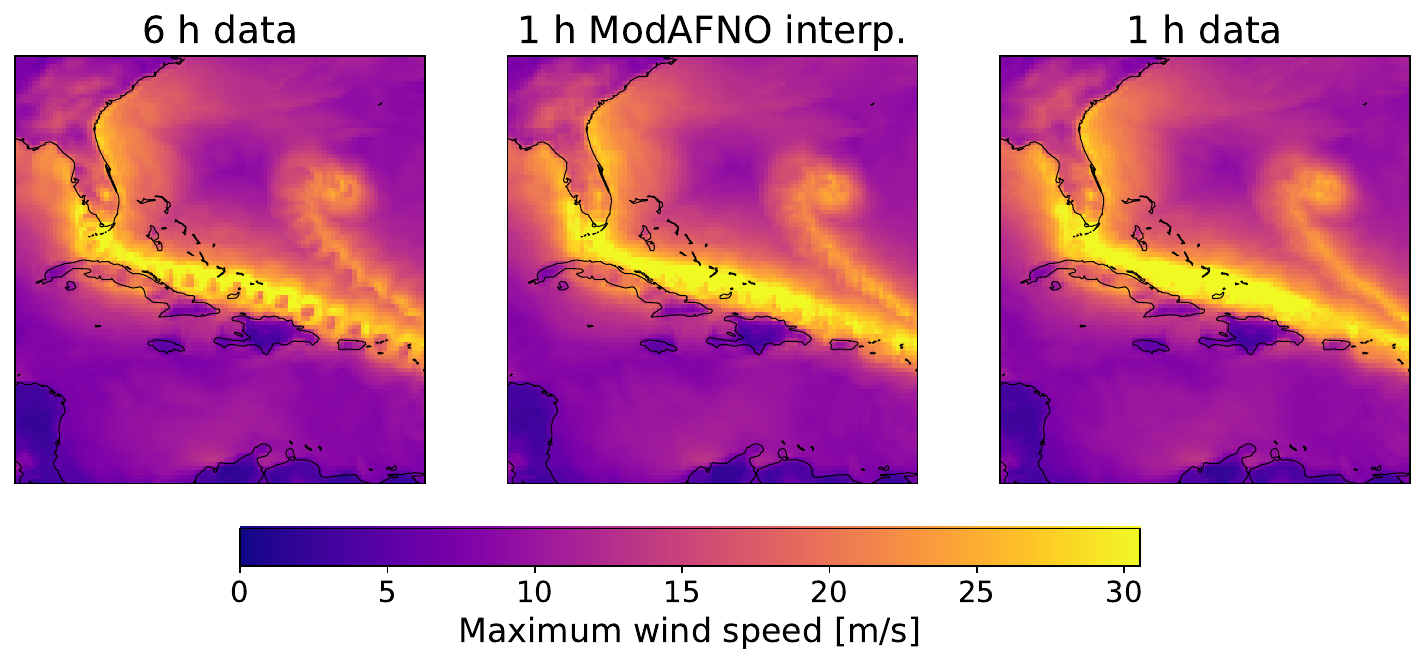}
  \caption{Maximum $10$~m wind speed during Hurricane Irma in the Caribbean and Florida on September 2--8, 2017. Left: obtained with $6$~h temporal resolution. Middle: with ModAFNO interpolation to $1$~h resolution. Right: the true maximum at $1$~h resolution.} \label{fig:max_U}
\end{figure}

To demonstrate the advantages of the $1$~h time interpolation, we reconstructed the wind speeds along the path of Hurricane Irma of September 2017, moving through the Caribbean towards Florida. In Fig.~\ref{fig:max_U}, we show the maximum wind speed, which is a good indicator of the level of wind-caused damage, computed from the original $6$~h resolution (left), interpolation of the $6$~h data to $1$~h (middle), and the actual $1$~h data (right). The $6$~h resolution data displays the maximum wind speeds fluctuating between weak and strong along the path, owing to the region of peak winds moving too fast for the $6$~h time resolution to capture it smoothly. The $1$~h interpolation captures the true maximum wind speeds much more accurately, even though some error remains particularly at the southern flank of the path of maximum winds.

\subsection{Extreme temperatures in a European heat wave}

\begin{figure}
  \centering
  \includegraphics[width=0.75\linewidth]{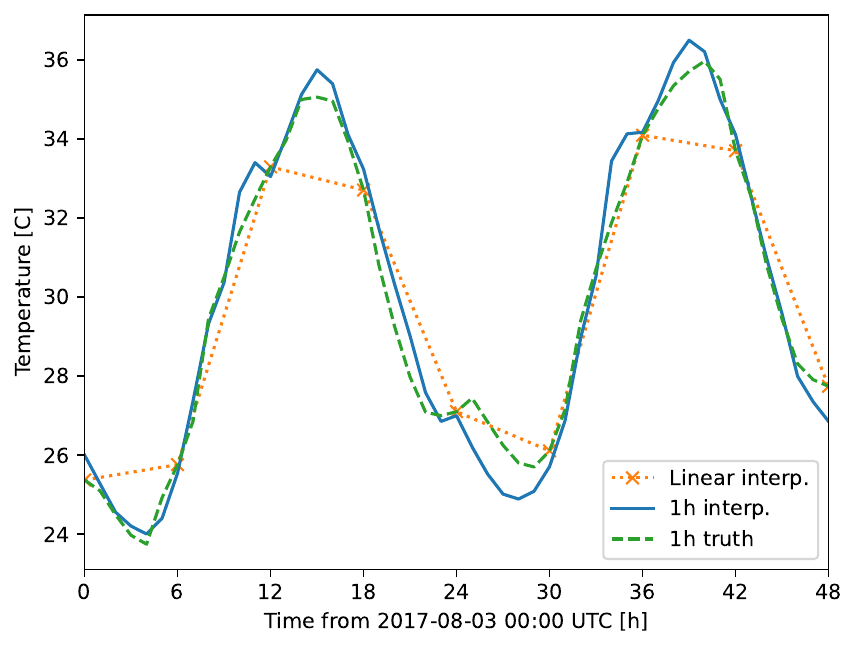}
  \caption{$2$ m temperature at the grid square containing the city of Milan, Italy, during a heat wave on August 3--4, 2017, as indicated by linear interpolation between $6$~h resolution time steps (orange dotted line), interpolation to $1$ h resolution (blue solid line) and the true $1$~h resolution data (green dashed line).} \label{fig:time_series_t2m}
\end{figure}

To demonstrate that the interpolation model has learned information about the physics of the atmosphere, we demonstrate another case of extreme weather prediction in the form of the temperatures during a European heatwave in August 2017. On August 3--4, the peak temperatures in Milan, Italy, occurred on both days approximately halfway between the $6$~h time steps given at 12 and 18~UTC. Thus, the $6$~h forecast would miss these extremes and underestimate the peak temperatures. In the plot shown in Fig.~\ref{fig:time_series_t2m}, we show that the $6$~h forecast linearly interpolated between time steps indeed predicts peak temperatures approximately $2\ \degree\mathrm{C}$ lower than the actual temperature. Meanwhile, the $1$~h interpolated forecast correctly infers that the hottest part of the day occurs between the endpoints of the interpolation period, and predicts the correct temperature with an accuracy of $0.5\ \degree\mathrm{C}$ (slightly overestimating on both days) and the timing of the highest temperature to within $1$~h.

\section{Discussion and summary} \label{sect:discussion}

A major advantage of ML weather forecast models over traditional NWP model is their orders of magnitude higher speed and hence cost efficiency. This advantage partially derives from the long time step used in ML weather models, such as $6$ hours in FourCastNet and GraphCast. However, this temporal resolution is insufficient for capturing quickly moving or rapidly evolving weather phenomena. Similar issues can arise when weather data is archived at a coarse temporal resolution due to storage constraints. To alleviate these issues, we have introduced a weather interpolation model that fills in the gaps between $6$~h time steps at $1$~h temporal resolution using a modulated AFNO layer that adapts to the desired target time step. We show that this model significantly outperforms linear interpolation at predicting the intermediate weather states. It also exhibits better skill than linear interpolation with extreme weather events such as fast-moving hurricanes and temperature maxima in heat waves.

Ideally, a model trained with a continuous input indicating the time step would generalize well enough to predict the weather state at any point between the endpoints of the interpolation spaced $6$~h apart. Thus, it should be possible to inference the model at arbitrarily high temporal resolution. Unfortunately, this turns out not to be the case with the model trained in this work: attempts to inference it at half-hour time intervals (e.g. $3.5$~h from the beginning) produced artifacts that render the prediction unusable. Thus, the model appears to have only properly learned the target times found in the training data, which are only available at integer hours. Generalizing the model training better to enable temporally continuous inference remains a challenge for further work.

Beyond the ability to resolve rapidly evolving weather events, we expect that ModAFNO and similar solutions are also applicable to the challenge of containing the ever-increasing data storage requirements of weather forecast and climate data. Availability of fast and accurate interpolation makes it feasible to store only limited amounts of data, using the interpolation to recover the intermediate time steps that were not stored, at a cost of a small amount of error due to the imperfect reconstruction. This capability will be particularly attractive if the model can be trained to generalize to arbitrary target times steps rather than just $1$~h time steps, as in this case it will enable the estimation of atmospheric states at arbitrary temporal resolution, which can be highly valuable in applications such as nowcasting.

The modulation concept in ModAFNO can also be reasonably expected to be adaptable to similar network architectures such as the spherical Fourier neural operator \citep{Bonev2023SFNO}. Furthermore, its use is not limited to interpolation and we expect it can be used to train weather forecast models with adjustable lead times. The embedding used to adjust ModAFNO can also be derived from other contexts besides lead time, allowing other types of conditional networks to be constructed.


\medskip
{
\small
\printbibliography
}

\end{document}